# Statistical handling of reproduction data for exposure-response modelling

Marie Laure Delignette-Muller[1,2*], Christelle Lopes[1], Philippe Veber[1] and Sandrine Charles[1,3]

[1] Université de Lyon, F-69000, Lyon; Université Lyon 1; CNRS, UMR5558, Laboratoire de Biométrie et Biologie Évolutive, F-69622, Villeurbanne, France

[2] Université de Lyon, F-69000, Lyon; VetAgro Sup Campus Vétérinaire de Lyon, F-69280 Marcy l'Etoile, France

[3] Institut Universitaire de France; 103 bd Saint-Michel; 75005 Paris, France.

## TOC/ Abstract

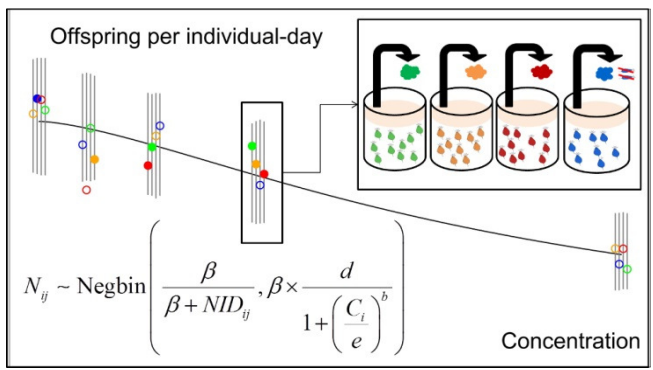

# Abstract

Reproduction data collected through standard bioassays are classically analyzed by regression in order to fit exposure-response curves and estimate $EC_x$ values (x% Effective Concentration). But regression is often misused on such data, ignoring statistical issues related to i) the special nature of reproduction data (count data), ii) a potential inter-replicate variability and iii) a possible concomitant mortality. This paper offers new insights in dealing with those issues. Concerning mortality, particular attention was paid not to waste any valuable data - by dropping all the replicates with mortality - or to bias $EC_x$ values. For that purpose we defined a new covariate summing the observation periods during which each individual contributes to the reproduction process. This covariate was then used to quantify reproduction - for each replicate at each concentration - as a number of offspring per individual-day. We formulated three exposure-response models differing by their stochastic part. Those models were fitted to four datasets and compared using a Bayesian framework. The individual-day unit proved to be a suitable approach to use all the available data and prevent bias in the estimation of $EC_x$ values. Furthermore, a non-classical negative-binomial model was shown to correctly describe the inter-replicate variability observed in the studied datasets.

# Introduction

During the last decade, many scientists[1–4] advocated the ban of the NOEC (No Observed Effect Concentration) and its replacement by the $EC_x$ (x% Effective Concentration). There is now a large consensus that the $EC_x$ has advantages over the NOEC and that it should be considered as an appropriate approach to quantify the effects of a contaminant on individual endpoints, provided that data are sufficient and properly analyzed to fit a exposure-response curve[5]. Nevertheless, some scientists severely criticize the way regression methods are sometimes misused (from a statistical point of view) to estimate $EC_x$[5]. Among their criticisms, Green *et al.*[5] stressed misuses concerning the nature of data (quantal, ordered score, count or continuous data) that is not always accounted for, the hierarchy of the experimental design (measurements made on multiple tanks of organisms or multiple organisms at each concentration) that is generally not considered, and the poor quality of some fits due to insufficient data. These criticisms are well founded and should not be neglected in exposure-response modelling. Within this context, new methods should be proposed to improve statistical handling of ecotoxicological data for estimating $EC_x$[6].

In the present work, we focused on the reproduction data, which are commonly used to estimate $EC_x$ from chronic toxicity tests. Reproduction data are count data, corresponding to the cumulated number of offspring at the end of the test[7]. Ordinary nonlinear regression is often performed to fit an exposure-response curve, even though data are unlikely to have a normally distributed error structure[8]. In 2008, the OECD guideline for *Daphnia magna* reproduction test[9] recommended weighted least squares to cope with a typical statistical problem encountered with count data: increased variance associated with increased observed number of offspring. In 2012, the revised version of the same guideline[7] stated the use of regression analysis to estimate $EC_x$, without mentioning any problem due to the nature of data, while recommending to transform data before performing statistical tests. As models and regression methods were specifically developed for count data, there is no reason to use least square methods falsely assuming a Gaussian distribution of the dependent variable[8,10]. For the analysis of count data from toxicity experiments the best distributions should recognize the discrete nature of the data (e.g. the Poisson distribution).

The use of the Poisson log-linear model had already been proposed for reproduction data[11,12]. Nevertheless, such a model suffers from the same drawbacks raised by Green *et al.*[5] concerning probit, logistic or Weibull models for survival data. Indeed, the replicate nature of the experiments is ignored and the potential overdispersion of data due to the inter-replicate variability cannot be properly described by the ordinary Poisson model, which assumes the variance of the response equal to its mean. Yet, overdispersion can be incorporated in a Poisson model in a variety of ways. Among them, quasi-Poisson and negative binomial regressions are the most popular[13–15]. As weighted least squares, quasi-Poisson regression does not strictly fall within the frame of maximum likelihood: there is generally no parametric distributional form of the model and so the likelihood cannot be expressed in closed form[10]. This complicates the use of some statistical tools such as information criteria based on the likelihood (AIC, BIC) classically used to compare models[15]. The negative binomial regression may be considered as more convenient as it is associated to a distributional form of a model which enables the expression of its likelihood. Another difference between quasi-Poisson and the classical negative binomial regression is discussed in Ver Hoef *et al.*[15]. In quasi-Poisson regression, the variance is a linear function of the mean, while in classical negative binomial regression the variance is a quadratic function of the mean. Such a difference may impact the fit of the model to data, and thus the parameter estimations, and the best choice between both regressions depends on the data. The classical form of the negative binomial distribution has already been used to describe overdispersion of ecotoxicological endpoints[16,17]. Nevertheless some authors have proposed non classical forms[18,19] that might be interesting to test on reproduction data, especially to describe a linear relation between variance and mean as in the quasi-Poisson regression[18].

Mortality is another problem often encountered in the analysis of reproduction data. OECD recommends first to record mortality of parent animals at least at the same timepoints that those at which offspring are counted, and second, when mortality occurs, to exclude the offspring of these parents from the calculations and to work with the total number of offspring produced per parent animal still alive at the end of the test[7]. When mortality is not negligible at the end of the test, notably for the highest concentrations, we may unfortunately lose valuable data by following this recommendation. In fact, parents may have reproduced before dying and we should not drop valuable information provided by the corresponding data, that certainly correspond to the response of the most sensitive animals. By dropping those data, we could even bias the results. Moreover, when mortality is high and when animals are not individually bred, data restricted to replicates without mortality at the end of the test may be insufficient to reasonably fit any exposure-response curve. In such cases, the total number of offspring is sometimes divided by the number of parents still alive at the end of the test without excluding replicates with mortality[20,21], what seems not satisfactory either, as it assigns the production of animals died before the end of the test to the only animals still alive at the end of the test. This biases the results by overestimating the production of each animal.

Wang *et al*[22] proposed to use a Poisson log-linear model and a logit model to describe reproduction data accounting for zeros in reproduction due to mortality. Nevertheless, they do not account for reduced but non-zero reproduction data due to mortality before the end of a toxicity test. Our idea is thus to define a new covariate from survival data, by summing the periods each parent is staying alive during the experiment. Such a covariate is classically used in epidemiology for incidence rate calculation: the number of observed cases is divided by the sum of periods for each individual during which the latter is at risk[23]. The incidence is then expressed as a number of cases per individual-day (or individual-month or –year). By analogy, the reproduction at each tested concentration can be characterized by the number of offspring per individual-day.

The aim of the present work is to propose a new way to analyse reproduction data for estimating $EC_x$, both taking into account mortality among parents without losing valuable data and describing potential inter-replicate variability using an appropriate statistical model.

## Materials and methods

**Data**

In order to present our new approach, we used four datasets corresponding to chronic laboratory bioassays on the freshwater invertebrate *Daphnia magna* exposed to four different contaminants: an organochlorine insecticide[24] and three metals, cadmium[21], copper[25] and zinc[25]. While the chlordan dataset was obtained from a classical experimental design with 10 animals per concentration held individually, the three other datasets correspond to different designs with animals held in groups of 10 or 20 (Table 1). We chose to include designs with animals held in groups, in particular because this type of design is required to achieve breeding for species such as hermaphroditic gastropods[26].

**Models**

During a chronic toxicity test, mortality among the parent animals is generally recorded daily, or at least at each timepoint the offspring are counted[7]. Using these survival data, it is thus possible to calculate the period during which each parent animal has stayed alive, or the period during which it may have reproduced. For such a calculation, an animal discovered dead at timepoint $t_{i+1}$ was assumed to be alive from the beginning of the experiment to time $\frac{t_{i+1} + t_i}{2}$. As commonly done in epidemiology for incidence rates we calculated, for each replicate (that may include more than one parent animal), the sum of the periods of observation of each animal before its death[23]. This sum was expressed as a number of individual-days. In the following, we will denote $N_{ij}$ the number of offspring in the replicate $j$ at the $i^{th}$ concentration $C_i$ and $NID_{ij}$ the number of individual-days for this replicate. The reproduction rate for a replicate can be expressed as the number of offspring per individual-day, that is $N_{ij}/NID_{ij}$. The calculation of $NID$ values was performed using the R software[27] (R code is provided in the Supporting Information).

To fit an exposure-response curve to data, it is necessary to choose a model characterized both by a deterministic part and a stochastic part. The latter is sometimes also called the error model. Concerning the deterministic part, we used the three parameter log-logistic model; it is widely used as it describes well a large number of exposure-response curves[28]. Notice that relevance of the derivation of the stochastic part would remain unchanged using any other deterministic part, such as log-normal or Weibull ones[28]. Thus we modeled the mean reproduction rate (in number of offspring per individual-day) at concentration $C_i$ with the following equation (eq 1).

$$f(C_i) = \frac{d}{1+\left(\frac{C_i}{e}\right)^b} \qquad (1)$$

where $d$ stands for the expected number of offspring per individual-day in the control, $e$ is the 50% effective concentration $EC_{50}$ and $b$ is a slope parameter.

In our work, we focused on the stochastic part in order to take into account the nature of reproduction data and the inter-replicate variability. For that purpose, we compared three different stochastic parts, all describing the number of offspring for concentration $i$ and replicate $j$ by a Poisson distribution of mean equal to the product of the number of individual-days $NID_{ij}$ by a term $f_{ij}$ differing between the three stochastic parts (eq. 2).

$$N_{ij} \sim \text{Poisson}\left(f_{ij} \times NID_{ij}\right) \qquad (2)$$

At first, we tested a simple model (called Poisson afterwards), by neglecting the potential inter-replicate variability, so assuming that $f_{ij}$ only depends on the concentration (eq. 3).

$$f_{ij} = f(C_i) \qquad (3)$$

For the two other stochastic parts, $f_{ij}$ was assumed to be variable from one replicate to another at a same concentration and to follow a gamma distribution. Gamma distributions denoted Gamma($\alpha, \beta$) are parameterized by the shape parameter $\alpha$ and the rate parameter $\beta$ all along this paper. In the following $\omega$ corresponds to an overdispersion parameter. In the second stochastic part, this gamma distribution was parameterized as in generalized linear models[14,29] with the shape parameter $\alpha$ fixed to $1/\omega$ (eq 4), while the third stochastic part was parameterized in a less classical way with the rate parameter $\beta$ fixed to $1/\omega$ (eq 5).

$$f_{ij} \sim \text{Gamma}\left(\frac{1}{\omega}, \frac{1}{\omega \times f(C_i)}\right) \quad (4)$$

$$f_{ij} \sim \text{Gamma}\left(\frac{f(C_i)}{\omega}, \frac{1}{\omega}\right) \quad (5)$$

Let us recall a general result[30]: if the distribution of y conditionally on $\theta$ is $y \sim \text{Poisson}(\lambda \times \theta)$ with $\lambda \sim \text{Gamma}(\alpha, \beta)$, then $y \sim \text{Negbin}(p, r)$ with $p = \frac{\beta}{\beta + \theta}$ and $r = \alpha$. Applying this general result to eq 2 mixed with eq 4 or eq 5, one can show that the second and third models can respectively be written as negative binomial models (eq 6 and eq 7 respectively):

$$N_{ij} \sim \text{Negbin}\left(\frac{1}{1 + \omega \times f(C_i) \times NID_{ij}}, \frac{1}{\omega}\right) \quad (6)$$

$$N_{ij} \sim \text{Negbin}\left(\frac{1}{1 + \omega \times NID_{ij}}, \frac{f(C_i)}{\omega}\right) \quad (7)$$

In the second model (eq 6), we get $\mu = f(C_i) \times NID_{ij}$ for the mean of $N_{ij}$, and $\sigma^2 = \mu(1 + \omega\mu)$ for its variance. The negative binomial distribution defined by Equation (6) thus assumes a quadratic mean-variance relationship. This assumption is the most classically made when using the negative binomial regression[10,15,17,30]. As done in Cameron and Trivedi[10], we call this model NegBin2 (eq 6), the "2" referring to the quadratic mean-variance relationship.

In the third model (eq 7), the mean is the same as in the NegBin2 model, but the mean-variance relationship is linear: $\sigma^2 = \mu(1 + \omega NID_{ij})$. Hence, the less classical parameterization of this model leads to the same assumption as in the quasi-Poisson regression, but preserves the complete formulation of the model likelihood. Here again, following the notation proposed by Cameron and Trivedi[10], we will call this model NegBin1 (eq 7), the "1" referring to the linear mean-variance relationship.

**Bayesian inference**

The three stochastic parts, Poisson (eq 3), NegBin1 (eq 7) and NegBin2 (eq 6), coupled with the three parameter log-logistic model (deterministic part), can be fitted to data using maximum likelihood or Bayesian inference. To our knowledge, there is no R package directly enabling the fit of the NegBin1 model by maximum likelihood but few R lines suffice to implement it using the mle function of the stats4 R package. It is also straightforward to fit the three models within a Bayesian framework that enables not only to easily calculate uncertainty on any function of the model parameters, but also to perform posterior predictive check from posterior distributions. This latter approach is particularly interesting to validate stochastic parts of models, which seemed of special interest in this work to compare the Poisson, NegBin1 and NegBin2 models.

Bayesian inference requires the definition of prior distributions on each model parameter. A prior distribution should describe the state of knowledge on each parameter before the experiment. The Poisson, NegBin1 and NegBin2 models based on eq. 1 as deterministic part, have three parameters ($d$, $e$, $b$) for the Poisson model or four parameters ($d$, $e$, $b$, $\omega$) for the NegBin1 and NegBin2 models. For parameters $d$ and $e$, a plausible range of values can generally be proposed before the experiment, from biological knowledge of the studied species and/or from previous experiments on the studied species with the same contaminant. Hence, for parameter $d$, the maximum number of offspring per individual-day of *Daphnia magna* in control, a prior uniform distribution between 0 and 20 was used (a case of 8 clutches during a 21-day test with an exceptionally high number of 60 eggs per clutch would give an extreme number of roughly 20 eggs per individual-day). For parameter $e$, i.e. the EC$_{50}$, we assumed that the experimental design was defined from a prior knowledge on the $EC_{50}$, with a probability of 95% for the expected value to lie between the smallest and the highest tested concentrations. Hence, a lognormal distribution for $e$ was calibrated from that prior knowledge. For the other parameters, noninformative distributions were chosen. Priors on the overdispersion parameter $\omega$ of the NegBin1 and NegBin2 models were described by log-uniform distributions between -4 and 4 in decimal logarithm. Finally, the slope parameter of the log-logistic model was characterized by a prior log-uniform distribution between -2 and 2 in decimal logarithm.

For each model, Monte Carlo Markov-Chain (MCMC) techniques were used to estimate the full joint posterior distribution of parameters from prior distributions and data. Computations were performed using the JAGS software via the R package rjags[31] (an R script with the example of the NegBin1 model is provided in Supporting Information). For each model and each dataset, three independent MCMC chains were run in parallel. From a short pilot run of the chains (5000 iterations after a burn-in phase of 5000 iterations too), the method proposed by Raftery and Lewis[32] (implemented in the raftery.diag function within the rjags[31] package) was used to calculate the number of iterations and the thinning required to accurately estimate the parameter quantiles. The three chains were then run using this required number of iterations and thinning. The convergence was checked again by displaying MCMC chain traces and by computing the Gelman and Rubin's statistics as modified by Brooks and Gelman[33]. For each parameter, its point estimate was defined as the median of its marginal posterior distribution, and the 95% credible interval was defined from the 2.5 and 97.5 percentiles of this distribution. Posterior predictive check was performed, especially to validate the stochastic part of each model. For that purpose, 5000 predicted values of each data point ($N_{ij}$ values) were calculated from the joint posterior distribution and covariates $C_i$ and $NID_{ij}$. For each data point, 95% (resp. 50%) prediction intervals were then calculated, from 2.5 and 97.5 percentiles (resp. 25 and 75 percentiles) of the predicted values. For each dataset, the predicted numbers of offspring $N_{ij}$ (point estimates defined as medians of predicted values and 95% prediction intervals) were graphically compared to the observed values. The coverage of 95%

(resp. 50%) prediction intervals was also calculated as the percentage of 95% (resp. 50%) prediction intervals encompassing the corresponding observed data point. Coverages of 95% and 50% prediction intervals are expected to be of 95% and 50% if the model gives a good description of the data, both considering its deterministic and stochastic parts. To compare the goodness-of-fit of our three models, we also recorded the deviance information criterion (DIC)[34]. In Bayesian inference, the minimum DIC highlights the model that best describes the data. In the same spirit as Akaike's criterion (AIC), the DIC is a goodness-of-fit criterion penalized by the complexity of the model, but with a different calculation of the penalization that is well suited for hierarchical models[35].

The full joint posterior distribution of parameters can be used to quantify the uncertainty on any function of the parameters, in particular the exposure-response curve. So for each dataset, each model and 100 values of concentrations regularly spread within the range of tested concentrations, we used the joint posterior distribution of parameters to simulate 5000 values of $f_{ij}$, the number of offspring per individual-day for various replicates. For each concentration, we calculated 2.5, 50 and 97.5 percentiles of simulated values, from which we got a point estimate and a 95% credible interval. From these results we were able to plot each exposure-response relation as a curve surrounded by 95% credible limits.

## Results and discussion

**Reproduction data after calculation of number of individual-days**

Figure 1 shows the number of offspring per individual-day as function of concentration for the three datasets. We clearly see that the fit of an exposure-response curve would not have been possible if we had excluded all the replicates with mortality (open circles in Figure 1), except in the case of the chlordan dataset. Working on the number of offspring per individual-day thus proved to be relevant in case of parent mortality before the end of the experiment. This approach is of special interest when animals are not individually bred, as the death of only one parent in a replicate would imply the exclusion of the whole replicate following OECD recommendations[7].

**Comparison of models**

The MCMC algorithm reached convergence for the three models on the four datasets, at a speed differing from one model to another. A greater number of iterations was always required to reach convergence with the NegBin2 model (from 22,000 to 49,000 iterations) than with the Poisson model (from 15,000 to 19,000 iterations) or the NegBin1 model (from 15,000 to 22,000 iterations). For each dataset and each model, the point estimates and credible intervals of each parameter are reported in Table 2. Exposure-response curves of the three models, surrounded by 95% credible limits, are reported in Figure 1. Figure 2 shows the predicted numbers of offspring $N_{ij}$ (point estimates and 95% prediction intervals) against observed $N_{ij}$ values for the three models and the four datasets, using colors to distinguish intervals containing the corresponding observed value from the others. Coverages of 95% and 50% prediction intervals were also reported in Table 3, together with DIC values.

The three fits gave fairly similar exposure-response curves (Figure 1) and parameter point estimates (Table 2). Differences appear between the three models while looking at the way each one described inter-replicate variability (Figure 1 and 2). The simplest model, the Poisson model, is clearly not able to describe the observed dispersion of data points (Figure 2). Depending on the dataset, 50 to 80% of the supposedly 95% prediction intervals encompass the corresponding observed value, for a 95% expected value. From coverage values of prediction intervals (Table 3), the NegBin1 and NegBin2 models seem to better describe the inter-replicate variability with coverage values of 95% prediction intervals from 95 to 100%. Nevertheless, while looking carefully at Figure 2, we notice that the Negbin2 models tends to overestimate the inter-replicate variability with some datasets, especially at low concentrations for chlordan and zinc. Indeed, for control data points, the length of prediction intervals is always greater with the NegBin2 model than with the NegBin1 model, with mean ratios between NegBin2 and NegBin1 interval lengths of 1.25, 1.06, 1.09 and 3.96 for chlordan, cadmium, copper and zinc, respectively. Let us recall that the NegBin2 model (eq 6) assumes a quadratic mean-variance relationship, while the NegBin1 model (eq 7) assumes a linear mean-variance relationship. This second hypothesis seems more relevant at least on chlordan and zinc datasets. For these two datasets, the non-negligible variability between low observed values (at high concentrations) together with the quadratic mean-variance relationship of the NegBin2 model could explain the overestimation of prediction intervals at low concentrations. Comparison of DIC values reinforces this conclusion (Table 3). The Poisson model appears as the worse model for the four datasets. Concerning negative binomial models, while the NegBin1 and NegBin2 models gave similar DIC values for the cadmium and copper datasets, the NegBin1 model gave smaller DIC values for the chlordan dataset and even more clearly for the zinc dataset.

Based on the previous results obtained on the four datasets, the NegBin1 model appears as a relevant candidate to take into account inter-replicate variability on reproduction data in exposure-response modelling when it is necessary. For a use on other datasets, we recommend a parallel fit of the Poisson and NegBin1 models. Their comparison using an information criterion such as the DIC may help to choose the most appropriate model. Indeed, it might not be necessary to take into account overdispersion for all datasets. When the Poisson model is sufficient to describe the dispersion of data points, it should be preferred. While the impact of the stochastic part of the fitted model on point estimates of parameters seems negligible, its impact on credible intervals is greater (Table 2). The Poisson model tends to underestimate the uncertainty with smaller credible intervals on each parameter, including the $EC_{50}$.

**Impact of statistical handling of data on $EC_x$ values**

Calculating the sum of the observation periods for each animal before its death reveals to be a good way to include replicates with mortality, from which data may be essential to fit an exposure-response curve. We believe this approach is far more reasonable than dividing the total number of offspring by the number of parents alive at the end of the test. Indeed, this would bias the observed reproduction in case of high mortality, by assigning the production of animals dying during the experiment to the only animals still alive at the end. Such a method is thus expected to underestimate the effect of the contaminant and so to overestimate $EC_x$ values. In order to evaluate the impact of such a bias on $EC_x$ estimates, we also implemented this classical method, called in the following the "per-alive" method, on the four datasets studied in the present work. Using the R package drc[36] we fitted the three parameter log-logistic model (eq 1) on the total number of offspring divided by the number of parents alive at the end of the test using a Gaussian stochastic part and without excluding replicates with mortality. $EC_{10}$, $EC_{20}$ and $EC_{50}$ values with their 95% confidence intervals given by the R package drc are reported in Figure 3. The same figure also shows $EC_x$ values, with 95% credible intervals, obtained with the three proposed models, Poisson, NegBin1 and NegBin2 fitted as previously described.

We can thus see that the choice between the three models (Poisson, NegBin1 and NegBin2) mainly impacts the length of credible intervals around $EC_x$ values, while the "per-alive" method can lead to highly biased $EC_x$ point estimates, more often with an unsafe overestimation of $EC_x$ (9/12 cases). We also notice that 95% confidence intervals given by the *drc* package in this context are unrealistic in half of the cases, with a negative lower bound. This confirms that even on the transformed variable defined by dividing the total number of offspring by the number of parents alive at the end of the test, the Gaussian model is not correct. It is thus far better to take into the discrete nature of data.

**Perspectives**

In the present work, we compared different stochastic models to take into account the inter-replicate variability within exposure-response modelling. A good description of variability and uncertainty is of crucial interest in a risk assessment context[37]. We showed that the non-classical NegBin1 model was a good candidate to account for overdispersion in reproduction data and thus to avoid an underestimation of uncertainty on parameter estimates with the Poisson model. Within a Bayesian framework, the full joint posterior distribution of parameters can be used to quantify the uncertainty on any risk indicator used in decision-making. Within Monte Carlo simulations, Bayesian network can be used for example to predict the effect of a realistic exposure, taking into account its own variability and uncertainty.

Another novelty in our approach is to use survival data at each observation time in order to define the number of offspring per individual-day. Nevertheless, concerning reproduction data, we only used the final observation that is the total number of offspring at the end of the test. Our goal was to improve the way data from reproduction tests are analyzed, especially to estimate $EC_x$ values, but also to cope with some statistical problems highlighted by others[5]. We could go further in our investigations, by building a global model simultaneously describing reproduction and survival data as functions of time. In such a model the number of individual-days ($NID$) could be more appropriately described as a function of time and concentration, not only as a covariate. Hierarchical models based on the Poisson log-linear model were recently proposed to take into account inter-individual and inter-brood variability, but without taking into account mortality within parents[38]. A more complete toxicokinetics/toxicodynamics model could be developed based on DEBtox models (Dynamic Energy Budgets applied to toxicity data)[25,39,21,40,41] or on another simpler mechanistic approach. While using mechanistic models, we are faced with the same issue: inter-replicate and/or inter-individual variability. Other authors have already identified the investigation of inter-individual variability as one of the most important areas for further research with DEB models[42]. Even though a first published work attempted to show that inter-individual variability could be important[43], statistical and modeling developments could still be improved to take into account the inter-individual variability within a toxicokinetics/toxicodynamics modeling approach. The combination of negative-binomial models for reproduction data with beta-binomial models for survival data[44] could be an interesting way to explore for that purpose.

## Associated content

### Supporting information

Two R scripts with the chlordan dataset are available free of charge via the Internet at http://pubs.acs.org. The first script allows calculation of NID values (Number of Individual-Days) from raw data. The second one applies gives an example of fit with the NegBin1 model.

## Author information

### Corresponding Author

* Tel.: +33-4-72-43-29-00, fax.: +33-4-72-43-13-88, e-mail: marielaure.delignettemuller@vetagro-sup.fr

## Acknowledgments

We would like to thank Martyn Plummer for developing the useful JAGS and r-jags tools and Cécile Poix for revising the English.

## References

(1) Crump, K. S. Calculation of Benchmark Doses from Continuous Data. *Risk Anal.* **1995**, *15*, 79–89.

(2) Jager, T. Some Good Reasons to Ban ECx and Related Concepts. *Environ. Sci. Technol.* **2011**, *45*, 8180–8181.

(3) Landis, W. G.; Chapman, P. M. Well Past Time to Stop Using NOELs and LOELs. *Integr. Environ. Assess. Manag.* **2011**, *7*, vi–viii.

(4) Fox, D. R. Response to Landis and Chapman (2011). *Integr. Environ. Assess. Manag.* **2012**, *8*, 1–4.

(5) Green, J. W.; Springer, T. A.; Staveley, J. P. The Drive to Ban the NOEC/LOEC in Favor of ECx Is Misguided and Misinformed. *Integr. Environ. Assess. Manag.* **2012**, 1–17.

(6) Fox, D. R.; Billoir, E.; Charles, S.; Delignette-Muller, M. L.; Lopes, C. What to Do with NOECS/NOELS--Prohibition or Innovation? *Integr. Environ. Assess. Manag.* **2012**, *8*, 764–766.

(7) OECD *OECD Guidelines for Testing of Chemicals 211. Daphnia Magna Reproduction Test*; Paris, France, 2012; pp. 1–25.

(8) O'Hara, R. B.; Kotze, D. J. Do Not Log-Transform Count Data. *Methods Ecol. Evol.* **2010**, *1*, 118–122.

(9) OECD *OECD Guidelines for Testing of Chemicals 211. Daphnia Magna Reproduction Test*; Paris, France, 2008; pp. 1–23.

(10) Cameron, A. C.; Trivedi, P. K. *Regression Analysis of Count Data*; Cambridge University Press, 2013; pp. 1–587.

(11) Bailer, A. J.; Oris, J. T. Modeling Reproductive Toxicity in Ceriodaphnia Tests. *Environ. Toxicol. Chem.* **1993**, *12*, 787–791.

(12) Zhang, J.; Bailer, a J.; Oris, J. T. Bayesian Approach to Estimating Reproductive Inhibition Potency in Aquatic Toxicity Testing. *Environ. Toxicol. Chem.* **2012**, *31*, 916–27.

(13) Zeileis, A.; Kleiber, C.; Jackman, S. Regression Models for Count Data in R. *J. Stat. Softw.* **2008**, *27*.

(14) Gelman, A.; Hill, J. *Data Analysis Using Regression and Multilevel/hierarchical Models*; Cambridge university press, 2007; pp. 111–116.

(15) Ver Hoef, J. M.; Boveng, P. L. Quasi-Poisson vs. Negative Binomial Regression: How Should We Model Overdispersed Count Data? *Ecology* **2007**, *88*, 2766–72.

(16) Noe, D. A.; Bailer, A. J.; Noble, R. B. Comparing Methods for Analyzing Overdispersed Count Data in Aquatic Toxicology. *Environ. Toxicol. Chem.* **2010**, *29*, 212–219.

(17) Bailer, A.; Elmore, R. Simulation Study of Characteristics of Statistical Estimators of Inhibition Concentration. *Environ. Toxicol. Chem.* **2000**, *19*, 3068–3073.

(18) Lindén, A.; Mäntyniemi, S. Using the Negative Binomial Distribution to Model Overdispersion in Ecological Count Data. *Ecology* **2011**, 1414–1421.

(19) Okamura, H.; Punt, A.; Amano, T. A Generalized Model for Overdispersed Count Data. *Popul. Ecol.* **2012**, 467–474.

(20) Ducrot, V.; Pery, A.; Mons, R.; Quéau, H.; Charles, S.; Garric, J. Dynamic Energy Budget as a Basis to Model Population Level Effects of Zinc Spiked Sediments in the Gastropod Valvata Piscinalis. *Environ. Toxicol. Chem.* **2007**, *26*, 1774–1783.

(21) Billoir, E.; Delhaye, H.; Forfait, C.; Clément, B.; Triffault-Bouchet, G.; Charles, S.; Delignette-Muller, M. L. Comparison of Bioassays with Different Exposure Time Patterns: The Added Value of Dynamic Modelling in Predictive Ecotoxicology. *Ecotoxicol. Environ. Saf.* **2012**, *75*, 80–6.

(22) Wang, S.; Smith, E. Adjusting for Mortality Effects in Chronic Toxicity Testing: Mixture Model Approach. *Environ. Toxicol. Chem.* **2000**, *19*, 204–209.

(23) Thrusfield, M. *Veterinary Epidemiology*; Second Ed.; John Wiley & Sons, 2013; pp. 37–49.

(24) Manar, R.; Bessi, H.; Vasseur, P. Reproductive Effects and Bioaccumulation of Chlordane in Daphnia Magna. *Environ. Toxicol. Chem.* **2009**, *28*, 2150–2159.

(25) Billoir, E.; Delignette-Muller, M. L.; Péry, A. R. R.; Charles, S. A Bayesian Approach to Analyzing Ecotoxicological Data. *Environ. Sci. Technol.* **2008**, *42*, 8978–8984.

(26) Ducrot, V.; Teixeira-Alves, M.; Lopes, C.; Delignette-Muller, M.-L.; Charles, S.; Lagadic, L. Development of Partial Life-Cycle Experiments to Assess the Effects of Endocrine Disruptors on the Freshwater Gastropod Lymnaea Stagnalis: A Case-Study with Vinclozolin. *Ecotoxicology* **2010**, *19*, 1312–1321.

(27) R Core Team R: A Language and Environment for Statistical Computing **2013**.

(28) Ritz, C. Toward a Unified Approach to Dose-Response Modeling in Ecotoxicology. *Environ. Toxicol. Chem.* **2010**, *29*, 220–229.

(29) Venables, W. N.; Ripley, B. D. *Modern Applied Statistics with S*; Springer, 2010; pp. 1–512.

(30) Gelman, A.; Carlin, J. B.; Stern, H. S.; Rubin, D. B. *Bayesian Data Analysis*; Chapman & Hall/CRC, 2003; pp. 1–668.

(31) Plummer, M. Rjags: Bayesian Graphical Models Using MCMC **2012**.

(32) Raftery, A.; Lewis, S. One Long Run with Diagnostics: Implementation Strategies for Markov Chain Monte Carlo. *Stat. Sci.* **1992**, *7*, 493–497.

(33) Brooks, S. P. B.; Gelman, A. G. General Methods for Monitoring Convergence of Iterative Simulations. **1998**, *7*, 434–455.

(34) Spiegelhalter, D. J.; Best, N. G.; Carlin, B. P.; van der Linde, A. Bayesian Measures of Model Complexity and Fit. *J. R. Stat. Soc. Ser. B* **2002**, *64*, 583–639.

(35) Ntzoufras, I. *Bayesian Modeling Using WinBUGS*; John Wiley & Sons: Hoboken, NJ, 2009; pp. 1–493.

(36) Ritz, C.; Streibig, J. C. Bioassay Analysis Using R. *J. Stat. Softw.* **2005**, *12*, 1–22.

(37) Cullen, A. C.; Frey, H. C. *Probabilistic Techniques in Exposure Assessment: A Handbook for Dealing with Variability and Uncertainty in Models and Inputs*; Springer: New York, NY, 1999; pp. 1–335.

(38) Zhang, J.; John Bailer, a.; Oris, J. T. Estimating Brood-Specific Reproductive Inhibition Potency in Aquatic Toxicity Testing. *Environmetrics* **2012**, *23*, 696–705.

(39) Kooijman, S.; Bedaux, J. Some Statistical Properties of Estimates of No-Effect Concentrations. *Water Res.* **1996**, *1354*, 1724–1728.

(40) Jager, T.; Vandenbrouck, T.; Baas, J.; De Coen, W. M.; Kooijman, S. a L. M. A Biology-Based Approach for Mixture Toxicity of Multiple Endpoints over the Life Cycle. *Ecotoxicology* **2010**, *19*, 351–361.

(41) Billoir, E.; Delhaye, H.; Clément, B.; Delignette-Muller, M. L.; Charles, S. Bayesian Modelling of Daphnid Responses to Time-Varying Cadmium Exposure in Laboratory Aquatic Microcosms. *Ecotoxicol. Environ. Saf.* **2011**, *74*, 693–702.

(42) Jager, T.; Barsi, A.; Hamda, N.; Martin, B.; Zimmer, E.; Ducrot, V. Dynamic Energy Budgets in Population Ecotoxicology: Applications and Outlook. *Ecol. Modell.* **2013**.

(43) Jager, T. All Individuals Are Not Created Equal; Accounting for Interindividual Variation in Fitting Life-History Responses to Toxicants. *Environ. Sci. Technol.* **2013**, *47*, 1664–1669.

(44) Price, W. J.; Shafii, B.; Seefeldt, S. S. Estimation of Dose–Response Models for Discrete and Continuous Data in Weed Science. *Weed Technol.* **2012**, *26*, 587–601.

Table 1. Experimental design for each dataset.

| Contaminant | Tested concentrations | Number of replicates per concentration | Number of animals per replicate | Number of measures = timepoints from 0 to 21 days |
|---|---|---|---|---|
| Chlordan ($\mu g.L^{-1}$) | 0, 0.18, 0.73, 1.82, 2.9, 7 | 10 | 1 | 21 |
| Cadmium ($\mu g.L^{-1}$) | 0, 0.37, 0.86, 1.64, 4.36 | 4 | 10 | 9 |
| Copper ($\mu g.L^{-1}$) | 0, 1.25, 2.5, 5, 10 | 3 | 20 | 15 |
| Zinc ($mg.L^{-1}$) | 0, 0.074, 0.22, 0.66 | 3 | 20 | 14 |

Table 2. Estimates of model parameters with 95% credible intervals in brackets.

| Parameter | Chlordan | Cadmium | Copper | Zinc |
|---|---|---|---|---|
| **Poisson model** | | | | |
| $d$ | 5.55 [5.27, 5.85] | 6.43 [6.27, 6.59] | 4.85 [4.78, 4.93] | 3.17 [3.10, 3.24] |
| $e$ (= $EC_{50}$) | 1.75 [1.47, 2.07] | 3.68 [3.48, 3.88] | 7.52 [7.30, 7.76] | 0.35 [0.33, 0.37] |
| $\log_{10} b$ | -0.03 [-0.09, 0.03] | 0.21 [0.15, 0.27] | 0.73 [0.69, 0.78] | 0.58 [0.55, 0.61] |
| **NegBin1 model** | | | | |
| $d$ | 5.60 [5.10, 6.20] | 6.52 [5.90, 7.40] | 4.85 [4.61, 5.11] | 3.17 [2.84, 3.57] |
| $e$ (= $EC_{50}$) | 1.66 [1.16, 2.25] | 3.59 [2.68, 4.41] | 7.58 [6.95, 9.37] | 0.35 [0.28, 0.56] |
| $\log_{10} b$ | -0.05 [-0.16, 0.06] | 0.18 [-0.07, 0.51] | 0.75 [0.63, 1.38] | 0.63 [0.47, 1.22] |
| $\log_{10} \omega$ | -0.91 [-1.17, -0.67] | -1.15 [-1.45, -0.79] | -1.69 [-2.10, -1.23] | -1.28 [-1.68, -0.79] |
| **NegBin2 model** | | | | |
| $d$ | 5.62 [5.02, 6.35] | 6.50 [5.91, 7.37] | 4.84 [4.59, 5.12] | 3.39 [2.44, 5.53] |
| $e$ (= $EC_{50}$) | 1.66 [1.10, 2.32] | 3.57 [2.71, 4.28] | 7.59 [6.95, 9.44] | 0.32 [0.17, 0.58] |
| $\log_{10} b$ | -0.06 [-0.16, 0.05] | 0.20 [-0.02, 0.51] | 0.74 [0.63, 1.42] | 0.53 [0.23, 1.25] |
| $\log_{10} \omega$ | -1.41 [-1.69, -1.15] | -1.89 [-2.19, -1.54] | -2.33 [-2.79, -1.81] | -0.70 [-1.11, -0.22] |

Table 3. DIC values and coverages of 95% and 50% prediction intervals for the three models fitted on the four datasets.

|  | Coverages of 95% (resp. 50%) prediction intervals in % | | | DIC values | | |
|---|---|---|---|---|---|---|
|  | Poisson | NegBin1 | NegBin2 | Poisson | NegBin1 | NegBin2 |
| Chlordan | 80 (25) | 95 (52) | 97 (52) | 540 | 490 | 497 |
| Cadmium | 50 (15) | 95 (60) | 95 (60) | 412 | 254 | 252 |
| Copper | 60 (33) | 100 (53) | 100 (53) | 224 | 182 | 183 |
| Zinc | 58 (42) | 100 (75) | 100 (75) | 246 | 150 | 173 |

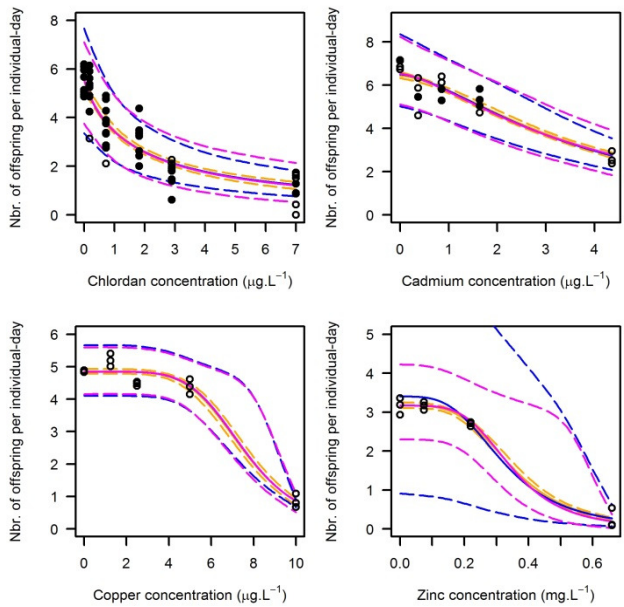

Figure 1. Reproduction data at the end of the experiment for the four studied datasets (chlordan, cadmium, copper and zinc) and fitted models. For each data point, the *y*-value corresponds to the total number of offspring collected from one replicate divided by the number of individual-days for this replicate. The data point is represented by an open circle if at least one parent died before the end of the experiment within the replicate, and by a solid circle otherwise. Fitted exposure-response curves (continuous lines) are also plotted surrounded by 95% credible limits (dotted lines) for the three models: Poisson (in orange), NegBin1 (in magenta) and NegBin 2 (in blue).

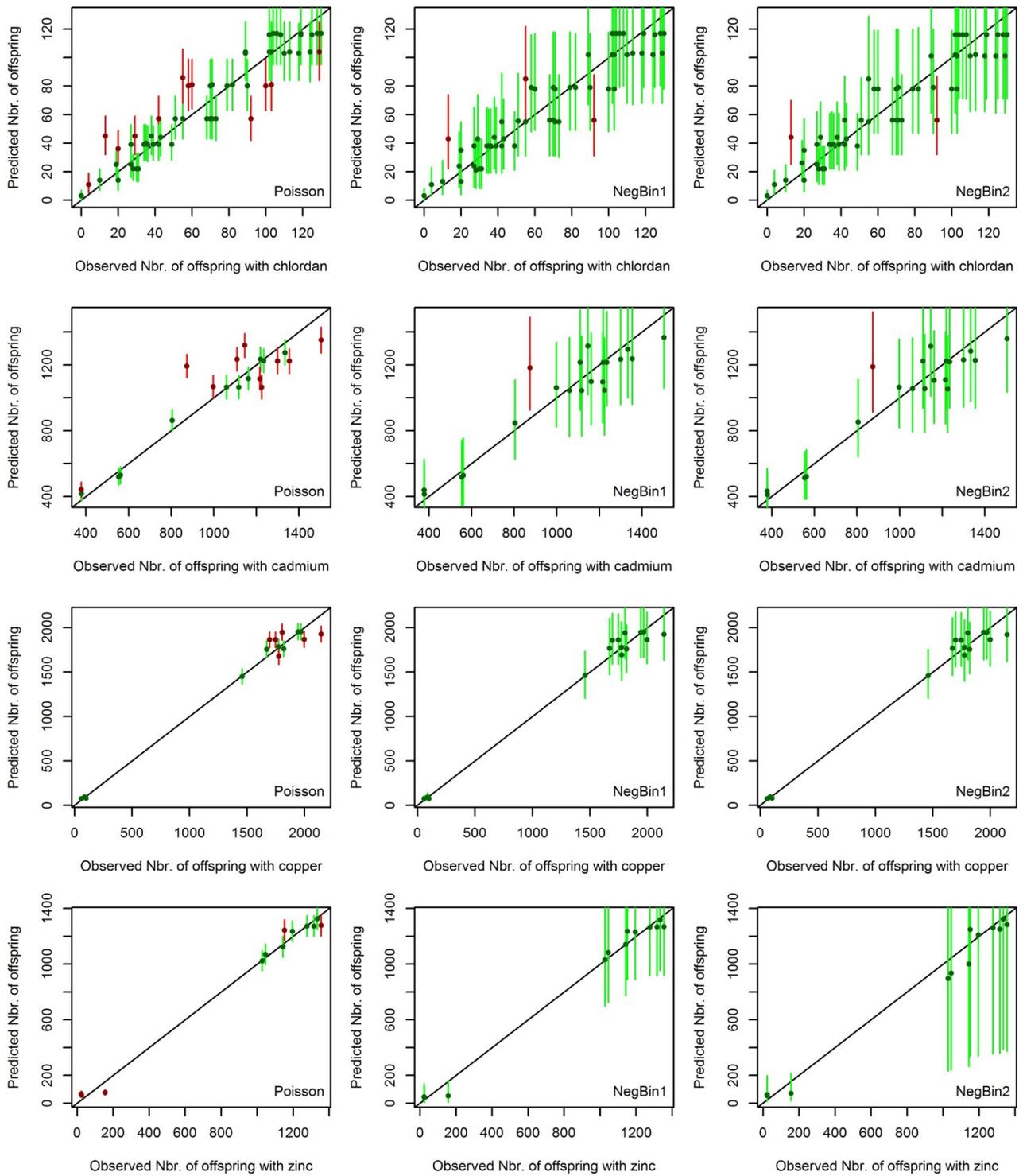

Figure 2. Plot of the predicted numbers of offspring (point estimates and 95% prediction intervals) against the observed numbers of offspring. Each panel line corresponds to one contaminant, each panel column to one model. The prediction intervals containing the corresponding observed value are colored in green and the others are colored in red

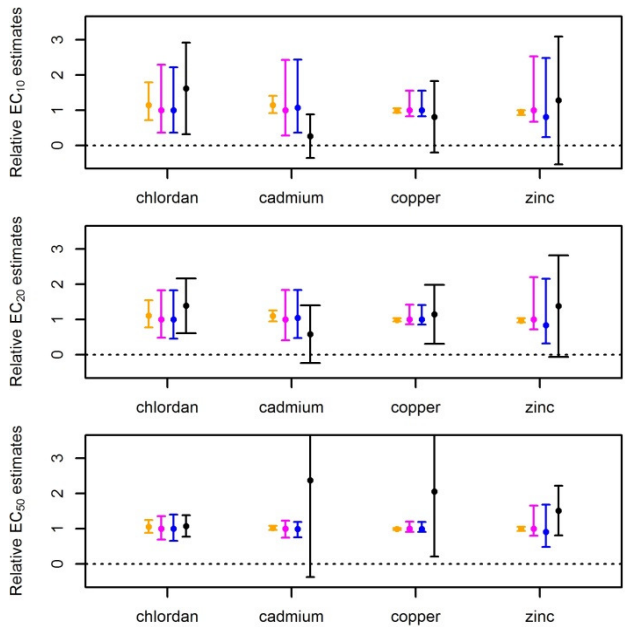

Figure 3. $EC_{10}$, $EC_{20}$ and $EC_{50}$ point estimates for each dataset and each method, with their 95% confidence or credible intervals depending on the type of inference (frequentist or Bayesian): Bayesian fit of Poisson (in orange), NegBin1 (in magenta), and NegBin2 (in blue) models and frequentist fit of a Gaussian model on the number of offspring divided by the number of parents alive at the end of the test (in black). For each dataset all $EC_x$ values were reported as relative values (all are divided by the point estimate from the NegBin1 model) because of the great variations in order of magnitude between datasets and between $EC_{10}$, $EC_{20}$ and $EC_{50}$ values.